\documentclass[aps,prb,showpacs,preprintnumbers]{revtex4-1}
\usepackage{mathrsfs}
\usepackage{amsmath}
\usepackage{amsfonts}
\usepackage{amssymb}
\usepackage{amsthm}
\usepackage{graphicx,natbib}
\usepackage{color}
\usepackage{hyperref}
\usepackage[caption=false]{subfig}
\providecommand{\U}[1]{\protect\rule{.1in}{.1in}}

\begin{document}
\title{Theoretical model for torque differential magnetometry of single domain magnets}
\author{Akashdeep Kamra$^{1,2}$}
\author{Michael Schreier$^{1}$}
\author{Hans Huebl$^{1,3}$}
\author{Sebastian T. B. Goennenwein$^{1,3}$}
\affiliation{$^{1}$Walther-Meissner-Institut, Bayerische Akademie der Wissenschaften, Walther-Meissner-Str. 8, D-85748 Garching, Germany}
\affiliation{$^{2}$Kavli Institute of NanoScience, Delft University of Technology,
Lorentzweg 1, 2628 CJ Delft, The Netherlands}
\affiliation{$^{3}$Nanosystems Initiative Munich (NIM), Schellingstr. 4, 80799 Munich, Germany}

\begin{abstract}
We present a generic theoretical model for {\it torque differential magnetometry} (TDM) - an experimental method for determining the magnetic properties of a magnetic specimen by recording the resonance frequency of a mechanical oscillator, on which the magnetic specimen has been mounted, as a function of the applied magnetic field. The effective stiffness change, and hence the resonance frequency shift, of the oscillator due to the magnetic torque on the specimen is calculated, treating the magnetic specimen as a single magnetic domain. Our model can deal with an arbitrary magnetic free energy density characterizing the specimen, as well as any relative orientation of the applied magnetic field, the specimen and the oscillator. Our calculations agree well with published experimental data. The theoretical model presented here allows to take full advantage of TDM as an efficient magnetometry method.

\end{abstract}

\maketitle


\section{Introduction}
While the exchange interaction is the pre-requisite for the existence of ferromagnetism in a solid, the static equilibrium and low energy dynamic properties of a magnet are determined by the dipolar and magneto-crystalline anisotropy. \cite{Kittel} Hence, several experimental methods, under the general name of ``magnetometry'', have been devised to determine the magnetic anisotropy of a given specimen. \cite{Cullity,Chikazumi} Some of these methods additionally allow investigating the saturation magnetization, magnetic switching, magnetic phase transitions and other properties of a magnet. \cite{Stipe,Nielsch,Okumura,Rathnayaka,Mueller,Li} State-of-the-art magnetometry is also sensitive enough for the investigation of thin magnetic films, which can have a strong shape (dipolar) or surface anisotropy. \cite{Bruno,Pinettes}

Torque magnetometry is a widely used magnetometry method and has been referred to as ``the most accurate means of measuring magnetic anisotropy''. \cite{Cullity,Chikazumi} In torque magnetometry, the mechanical torque exerted on a magnetic specimen by an externally applied magnetic field is recorded as a function of the field's orientation in a given plane of interest. Since torque can be expressed in terms of the derivative of the free energy density $F$, the experimental data can be used to infer the constants parameterizing $F$. \cite{Chikazumi} Cantilever torque magnetometry (CTM) \cite{Rossel} takes advantage of the small stiffness of AFM cantilevers to detect very small torques. The magnetization sensitivity of CTM is comparable to SQUID magnetometry \cite{Codjovi} over a broad temperature range. \cite{Rigue,Brugger,Slageren} An important advantage of torque magnetometry is its relatively fast response which allows for the investigation of dynamic phenomena in magnets and high $T_c$ superconductors.
\cite{Brugger,Schnack}

Instead of measuring the static ``magnetic force" (the DC torque) on a cantilever, one can also study the shift in the resonance frequency of the cantilever as a function of the applied magnetic field. The magnetic field dependence of the resonance frequency comes about via enhanced (or reduced) stiffness of the cantilever owing to the change in magnetic (in addition to elastic) energy as the cantilever deviates from equilibrium. Although this technique has simply been called ``cantilever magnetometry'' in the literature, \cite{Stipe,Weber,Buchter} it is more appropriate to call it {\it torque differential magnetometry} (TDM), to emphasize the fact that the derivative of torque, as opposed to the torque itself, is measured as will be discussed in Sec. \ref{Theory}. The relation between DC torque magnetometry and TDM is analogous to the relation between contact mode and frequency modulated AFM. \cite{Giessibl} TDM thus offers similar advantages - namely less $1/f$ noise, low drift and higher sensitivity at a 
given measurement rate.~\cite{Albrecht} Recent advances in using quartz tuning forks, instead of cantilever systems, for microscopy \cite{Giessibl,Rychen} and magnetometry \cite{Kamra,Todorovic} have made TDM particularly attractive due to the simplicity and wider operation range of the experimental setup. Hence, in this paper, we use the terms `cantilever' and `mechanical oscillator' (or simply `oscillator') interchangeably.

\begin{figure}[htb]
\centering
\subfloat[]{\includegraphics[width=8.5cm]{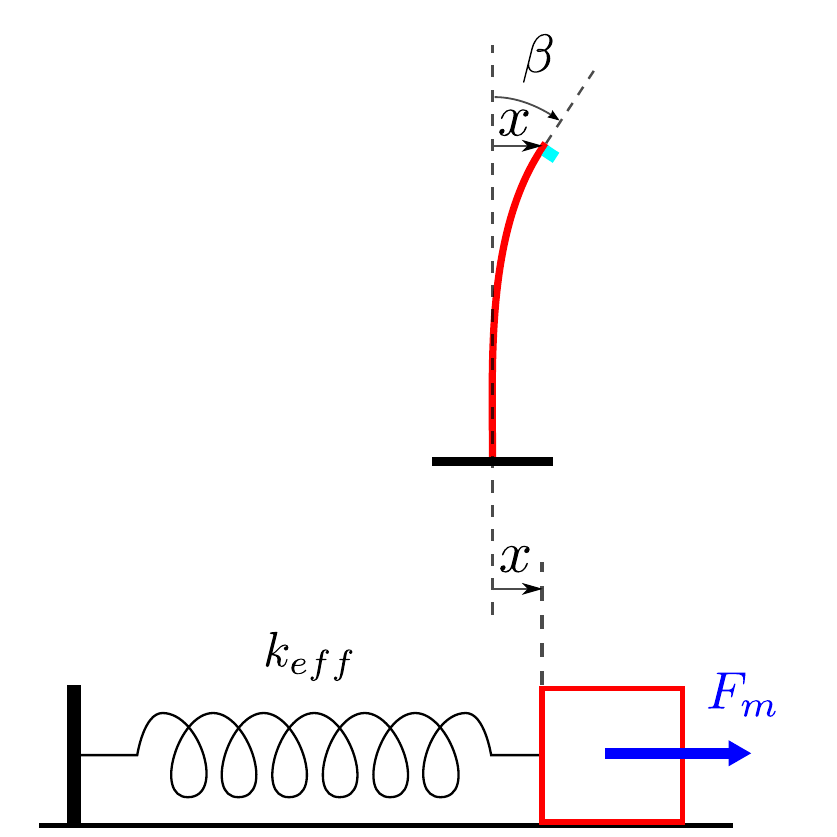}} \qquad
\subfloat[]{\includegraphics[width=8.5cm]{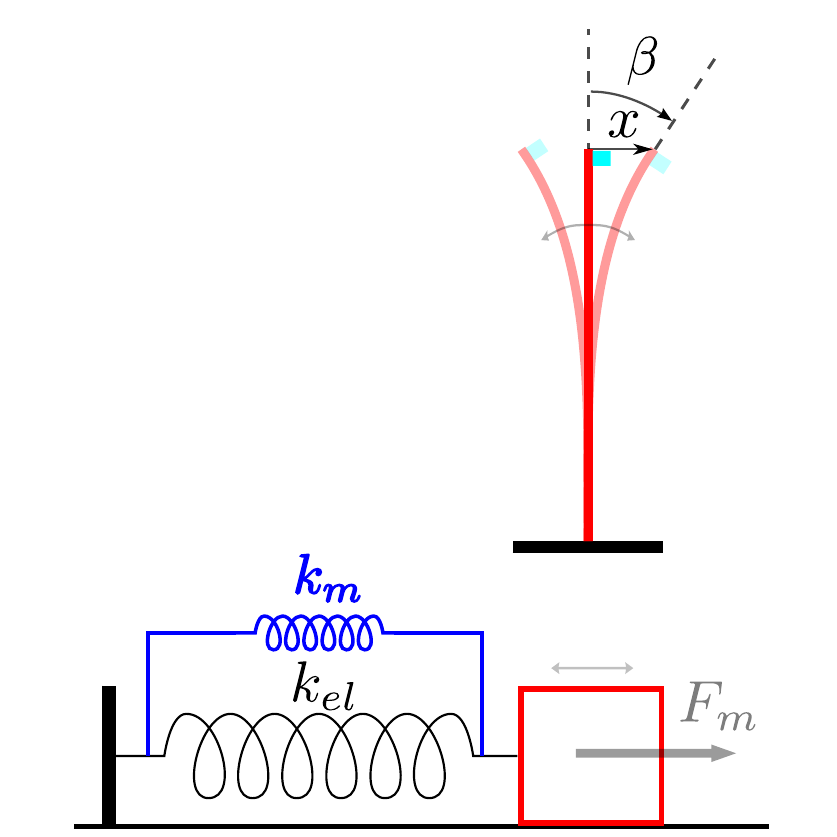}}
\caption{Comparison between (a) cantilever torque magnetometry (CTM) and (b) torque differential magnetometry (TDM). The magnetic specimen (light blue) is mounted at the tip of the cantilever (red), the motion of which can be modeled by an effective mass and spring system. (a) In CTM, one measures the static equilibrium deflection of the cantilever, which translates to the torque (modeled as an effective force $F_m$) exerted on the magnetic specimen by the applied magnetic field. (b) In TDM, one measures the magnetic field dependent resonance frequency shift of the cantilever which translates to a stiffness change due to magnetic torque $k_m$ [cf. Eq.(\ref{km1})]. In both (a) and (b), the physical quantity that the experiment measures is shown in blue color.}
\label{ctmvstdm}
\end{figure}

In contrast to torque magnetometry, \cite{Chikazumi} a consistent and complete theoretical modeling of TDM is still lacking in literature. Here, we present a generic formalism to calculate the magnetic field dependent shift in the resonance frequency of the mechanical oscillator, on which the magnetic specimen has been mounted, for any given magnetic free energy density, and configuration of the specimen, the oscillator and the applied magnetic field. We work within the macrospin approximation [Stoner-Wohlfarth (SW) model] \cite{SW} treating the specimen as a single domain magnet. The theoretical formalism is developed in Sec.~\ref{Theory}, followed by comparison of our model to existing literature in Sec.~\ref{MFD}. The high magnetic field limit, which is the normal mode of operation in conventional torque magnetometry, is discussed for TDM in Sec.~\ref{HFL}. Section~\ref{HFL} (and Appendix \ref{mfsdsec}) discusses some generic principles which can be employed in 
determining the required properties of the magnetic specimen in a simple and efficient manner. We conclude with a short discussion in Sec.~\ref{Conclusion}.

\section{Theory}\label{Theory}
We start our discussion of TDM by considering the properties of the mechanical oscillator to which the magnetic specimen is attached. Any mechanical oscillator can be modeled as an effective mass and spring system [Fig.~\ref{ctmvstdm}(a)] with $x$ denoting the displacement of its tip from the equilibrium position. \cite{Morse} The resonance frequency is then expressed in terms of effective mass ($m_{eff}$) and spring constant ($k_{eff}$)
\begin{equation}\label{ffull}
f = \frac{1}{2 \pi} \sqrt{\frac{k_{eff}}{m_{eff}}}.
\end{equation}
For a small displacement $x$ about the equilibrium point, the restoring force is given by the sum of elastic [$F_e(x)$] and magnetic [$F_m ( x)$] forces. $F_m (x)$ is an effective force representing the effect of the torque $\tau_m^{\perp} ( x)$ exerted on the magnetic specimen by the applied magnetic field. \cite{note1} The superscript $\perp$ denotes that the component of torque perpendicular to the plane of oscillation should be considered as discussed below in Sec. \ref{tauinqe}. Assuming an effective oscillator length $L_e$ (distance between the tip and an effective oscillation center), \cite{Morse} and Taylor expanding the torque $\tau_m^{\perp} (x)$ around the 
equilibrium position, we obtain the following for the restoring force:
\begin{eqnarray}
 F_r & = & - k_{el} x + \frac{1}{L_e} \left. \frac{d \tau_m^{\perp}}{d x} \right|_{eq} x \ = \ - k_{el} x - k_m(B_{ext}) x,
\end{eqnarray}
where $k_{el}$ is the effective elastic spring constant, and $|_{eq}$ denotes that the derivative has been calculated at the magnetization equilibrium configuration. Transforming the torque derivative from the linear variable $x$ to the angular variable $\beta$ ($ dx = L_e d\beta$) in the equation above, the effective spring constant due to magnetic torque becomes
\begin{eqnarray}\label{km1}
 k_m(B_{ext}) & = & - \frac{1}{L_e^2} \left. \frac{d \tau_m^{\perp}}{d \beta} \right|_{eq}.
\end{eqnarray}
Considering $k_{eff} = k_{el} + k_m(B_{ext})$ with $k_m(B_{ext}) \ll k_{el}$ in Eq. (\ref{ffull}), we introduce the frequency shift owing to the magnetic torque as $\Delta f/ f_{el} = \Delta k / 2 k_{el}$ which yields [with $\Delta k = k_m$ and Eq. (\ref{km1})]
\begin{equation}\label{deltaf}
\Delta f = f_{el} \frac{k_m}{2 k_{el}} = -  \frac{f_{el}}{2 k_{el} L_e^2} \left. \frac{d \tau_m^{\perp}}{d \beta} \right|_{eq}
\end{equation}
for the magnetic field dependent resonance frequency shift $\Delta f = f(B = B_{ext}) - f(B = 0)$. Thus, the frequency shift measured in a TDM experiment is proportional to the magnetic torque derivative. \cite{note2} In writing the above equation, we have disregarded any changes in the elastic properties of the magnetic specimen owing to magnetostriction. \cite{Chikazumi} 

\begin{figure}[htb]
 \centering
 \includegraphics[width=8.5cm]{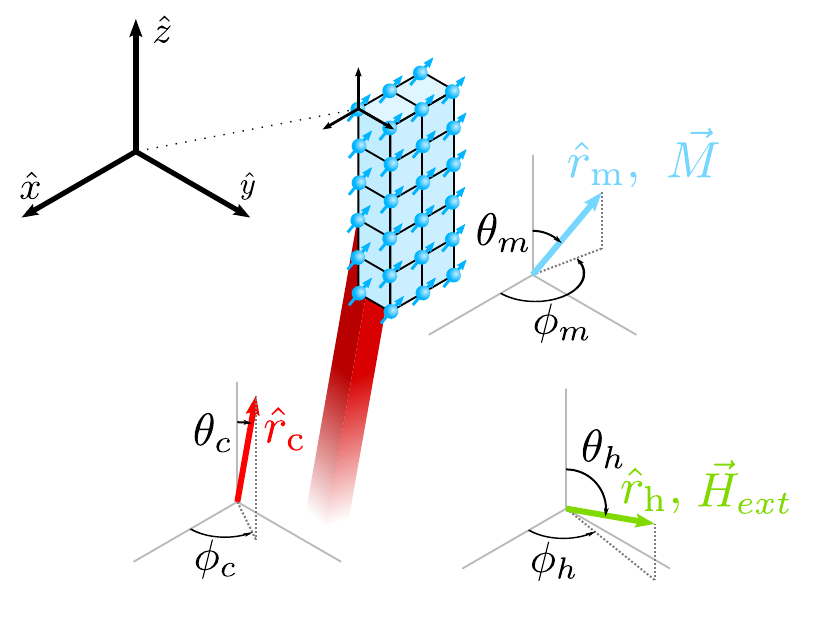}
 \caption{Schematic of the magnetic specimen (light blue) mounted on a cantilever (red) in an applied magnetic field. $\hat{\pmb{r}}_c$ (red arrow), $\hat{\pmb{r}}_m = \vec{M}/|M|$ (blue arrow) and $\hat{\pmb{r}}_h = \vec{H}/|H| $ (green arrow) denote the unit vectors along the oscillator axis, magnetization and applied magnetic field, respectively. The angles characterizing the unit vectors are depicted in the specimen frame of reference.}
 \label{angles}
\end{figure}

In the following subsection, we express the required derivative of the `perpendicular' component of the magnetic torque in terms of the magnetic free energy density and the variables defining the configuration of the system. Unless stated otherwise, we work in a spherical polar coordinate system attached to the lattice of the magnetic specimen. The relevant variables that characterize the system are summarized below (see Fig.~\ref{angles}).
\begin{table}[tb]
\begin{center}
\begin{tabular}{|l|p{8cm}|}
\hline
 $\theta_h$, $\phi_h$ & Instantaneous polar and azimuthal angles of the magnetic field direction. \\ \hline
 $\theta_{h0}$, $\phi_{h0}$ & Values of $\theta_h$ and $\phi_h$ for equilibrium orientation of the oscillator. \\ \hline
 $\theta_m$, $\phi_m$ & Instantaneous polar and azimuthal angles of the magnetization direction. \\ \hline
 $\theta_{m}^h$, $\phi_{m}^h$ & (Quasi) Equilibrium values of $\theta_m$ and $\phi_m$ for a given magnetic field. These are functions of the angles that define the magnetic field direction ($\theta_h, \phi_h$). \\ \hline
 $\theta_{m0}^h$, $\phi_{m0}^h$ & Values of $\theta_m^h$ and $\phi_m^h$ for the magnetic field orientation when the oscillator is in equilibrium position. This implies $\theta_{m0}^h = \theta_m^h(\theta_{h0},\phi_{h0})$ and $\phi_{m0}^h = \phi_m^h(\theta_{h0},\phi_{h0})$. \\ \hline
 $\theta_{c0}$, $\phi_{c0}$ & Equilibrium values of $\theta_c$ and $\phi_c$. The tip oscillates along the $\hat{\pmb{\theta}}_{c0}$ direction. \\
 \hline
\end{tabular}
\end{center}
\caption{Description of the polar coordinates used to specify the directions of the applied magnetic field, the magnetization and the oscillator. The frame of reference, with respect to which the angles above are defined, is attached to the magnetic specimen.}
\end{table}
Without loss of generality, we consider the motion of the oscillator tip to be along the $\hat{\pmb{\theta}}_{c0}$ direction. 

\subsection{Torque in (quasi) equilibrium}\label{tauinqe}

The magnetic free energy density (henceforth simply called `free energy density') is written as $F(M_s,\theta_{m},\phi_m,H_{ext},\theta_h,\phi_h)$, where $M_s$ is the saturation magnetization density of the specimen, and $H_{ext}$ is the magnitude of the applied magnetic field. Within the macrospin model, \cite{SW} we consider a uniformly magnetized sample which implies that the variables $\theta_{m}$ and $\phi_m$ are position independent. The effective magnetic field is given by
\begin{eqnarray}\label{heff}
 \mu_0 \pmb{H}_{eff} & = & - \pmb{\nabla}_M F = - \frac{\partial F}{\partial M_s} \hat{\pmb{r}}_m - \frac{1}{M_s}  \frac{\partial F}{\partial \theta_{m}} \hat{\pmb{\theta}}_m - \frac{1}{M_s \sin{\theta_{m}}}  \frac{\partial F}{\partial \phi_m} \hat{\pmb{\phi}}_m,
\end{eqnarray}
where we have used a spherical coordinate system i.e. $\pmb{M} = M_s (\hat{\pmb{r}}_m, \hat{\pmb{\theta}}_m, \hat{\pmb{\phi}}_m)$. The stable equilibrium values of $\theta_{m}$ and $\phi_m$ are obtained by minimizing the free energy [$\partial F/ \partial ( \theta_{m},\phi_m) = 0, (\partial^2F/\partial \theta_m^2) (\partial^2F/\partial \phi_m^2) - (\partial^2F/\partial \theta_m \partial \phi_m)^2 >0, \partial^2F/\partial \theta_m^2 > 0 $]. Let us call these values $\theta_m^h \equiv \theta_m^h(\theta_h,\phi_h)$ and $\phi_m^h \equiv \phi_m^h(\theta_h,\phi_h)$. Here the superscript $h$ emphasizes that these are the values for a given applied magnetic field magnitude and direction.

It is mathematically convenient to separate the free energy density $F = F^i + F^e$ into internal free energy density $F^i$ (consisting of anisotropy, magnetostatic energy etc.) and external free energy density $F^e = F_{Zeeman} = - \mu_0 \pmb{M} \cdot \pmb{H}_{ext}$. This separation along with the mathematical condition for equilibrium:
\begin{eqnarray}
  \left. \frac{\partial F}{\partial (\theta_{m},\phi_m)} \right|_{(\theta_m^h,\phi_m^h)} = \left. \frac{\partial F^e}{\partial (\theta_{m},\phi_m)} \right|_{(\theta_m^h,\phi_m^h)} + \left. \frac{\partial F^i}{\partial (\theta_{m},\phi_m)} \right|_{(\theta_m^h,\phi_m^h)} & = & 0,\label{intext}
\end{eqnarray}
allows us to express the component of the externally applied field orthogonal to the magnetization in terms of the derivatives of $F^i$ at magnetic equilibrium conditions, i.e.,
\begin{equation}\label{hextfi}
\pmb{r}_m^h \times \mu_0 \pmb{H}_{ext} = \pmb{r}_m^h \times -\pmb{\nabla}_M F^e|_{(\theta_m^h,\phi_m^h,\theta_h,\phi_h)} = \pmb{r}_m^h \times \pmb{\nabla}_M F^i|_{(\theta_m^h,\phi_m^h)} .
\end{equation}
The advantage of this substitution is that while $F^e$ is an explicit function of all four variables $\theta_m^h,\phi_m^h,\theta_h,\phi_h$, $F^i$ involves only the first two variables. This leads to simpler expressions in the rest of the analysis (c.f. Appendix A).

The total torque exerted by an external magnetic flux density $\pmb{B}_{ext} = \mu_0 \pmb{H}_{ext}$ on a magnetization distribution $\pmb{M}(\pmb{r})$ is given by \cite{Jackson}
\begin{eqnarray}
 \pmb{\tau}_{m} & = & \int_{\mathcal{V}} \pmb{M}(\pmb{r}) \times \pmb{B}_{ext}(\pmb{r}) \  d^3 r.
\end{eqnarray}
For the case of uniform magnetization and magnetic field, the torque experienced by the magnetic specimen in (quasi) equilibrium becomes
\begin{eqnarray}
 \pmb{\tau}_{m} & = & M_s V (\pmb{r}_m^h \times \mu_0 \pmb{H}_{ext}) ,
\end{eqnarray}
where $V$ is the volume of the magnetic specimen. Using Eqs. (\ref{heff}) and (\ref{hextfi}),
\begin{eqnarray}\label{torquegen}
 \pmb{\tau}_{m} & = &  V \left(  F^i_{\theta_{m}}(\theta_m^h,\phi_m^h) \ \hat{\pmb{\phi}}_m^h - \frac{1}{ \sin{\theta_m^h}}  F^i_{\phi_m}(\theta_m^h,\phi_m^h) \ \hat{\pmb{\theta}}_m^h \right),
\end{eqnarray}
where we adapt the compact notation $\partial F^i /\partial \theta_{m} |_{(\theta_m^h,\phi_m^h)} = F^i_{\theta_{m}}(\theta_m^h,\phi_m^h)$ and so on.

The motion of the oscillator tip and hence the restoring force is along the $\hat{\pmb{\theta}}_{c0}$ direction (Figs.~\ref{ctmvstdm} and~\ref{angles}). This implies that the relevant component of the torque (corresponding to effective force along $\hat{\pmb{\theta}}_{c0}$) is perpendicular to the plane of oscillation and points along $\hat{\pmb{\phi}}_{c0}$:
\begin{eqnarray}
 \tau_m^{\perp} & = & \pmb{\tau}_{m} \cdot \hat{\pmb{\phi}}_{c0}.
 \end{eqnarray}
 Using Eq. (\ref{torquegen}),
 \begin{eqnarray}
   \tau_m^{\perp} & = & V \left(  F^i_{\theta_m}(\theta_{m}^h,\phi_{m}^h) \ \hat{\pmb{\phi}}_{m}^h \cdot \hat{\pmb{\phi}}_{c0}  - \frac{1}{ \sin{\theta_{m}^h}}  F^i_{\phi_m}(\theta_{m}^h,\phi_{m}^h) \ \hat{\pmb{\theta}}_{m}^h \cdot \hat{\pmb{\phi}}_{c0} \right), \label{taur1} \\
   & = & V \left[   F^i_{\theta_m} \left(\theta_{m}^h,\phi_{m}^h\right) \ \cos{(\phi_{m}^h - \phi_{c0})}   -  F^i_{\phi_m}(\theta_{m}^h,\phi_{m}^h) \ \cot{(\theta_{m}^h)} \sin{(\phi_{m}^h - \phi_{c0})} \right]. \label{taur2} 
\end{eqnarray}
The scalar products $\hat{\pmb{\theta}}_{m}^h \cdot \hat{\pmb{\phi}}_{c0}$ and $\hat{\pmb{\phi}}_{m}^h \cdot \hat{\pmb{\phi}}_{c0}$ have been calculated in Appendix B. The expression obtained above is an explicit function of two variables ($\theta_m^h,\phi_m^h$) which are implicitly dependent on the magnetic field direction. 

Eq. (\ref{taur2}) enables us to obtain the torque, or equivalently the effective force along the deflection direction, experienced by the oscillator in a quasi-static state. Eq. (\ref{taur2}) thus represents a generic description of CTM measurements. In contrast, TDM measures the derivative of this torque with respect to the deflection angle, requiring a more sophisticated analysis.

\subsection{Oscillator deflection and torque derivative}

Before we proceed with the calculation of the torque derivative, let us first emphasize that the deviation angle $\beta$ [Fig.~\ref{ctmvstdm}(b)] enters the torque expression [Eq. (\ref{taur2})] via the magnetic field direction. A deflection of the oscillator (tip moves along $\hat{\pmb{\theta}}_{c0}$) from its equilibrium orientation by an angle $\beta = - \alpha$ mathematically implies that the lattice coordinate system has rotated about the axis parallel to $\hat{\pmb{\phi}}_{c0}$ and passing through the (effective) center of the oscillator, by the angle $-\alpha$. In the lattice coordinate system, this can be visualized as a rotation of the lab frame by an angle $+\alpha$. Since the magnetic field is fixed in the lab frame of reference, the net effect of this deflection is to rotate the magnetic field vector by an angle $+\alpha$ in the lattice frame of reference. We thus obtain the new direction of magnetic field in the lattice coordinate system as a function of $\alpha$.

The rotation operator written in Cartesian coordinate basis for a small rotation ($\alpha \ll 1$) about a unit vector $\hat{\pmb{u}} = u_x \hat{\pmb{x}} + u_y \hat{\pmb{y}} + u_z \hat{\pmb{z}} = [u_x ~ u_y ~ u_z]^T$ passing through the origin is given by \cite{Morse2}
\begin{eqnarray}
 \tilde{R}^\alpha(u_x,u_y,u_z)  & = & \left[ \begin{array}{ccc}
                                    1 & - \alpha u_z & \alpha u_y \\
                                    \alpha u_z & 1 & - \alpha u_x \\
                                    - \alpha u_y & \alpha u_x & 1 
                                   \end{array} \right].
\end{eqnarray}
For the case at hand, the unit vector $\hat{\pmb{\phi}}_{c0}$ is written as $[-\sin{\phi_{c0}} ~ \cos{\phi_{c0}} ~ 0]^T$ in Cartesian coordinates. The unit vector along the equilibrium magnetic field ($\hat{\pmb{h}}_0$) is then given by $[\sin{\theta_{h0}} \cos{\phi_{h0}} ~ \sin{\theta_{h0}} \sin{\phi_{h0}} ~ \cos{\theta_{h0}}]^T$. Therefore the rotated unit vector in Cartesian coordinates is given by
\begin{eqnarray}
 \hat{\pmb{h}}^\prime & = & \left[ \begin{array}{ccc}
                                    1 & 0 & \alpha \cos{\phi_{c0}} \\
                                    0 & 1 & \alpha \sin{\phi_{c0}} \\
                                    - \alpha \cos{\phi_{c0}} & - \alpha \sin{\phi_{c0}} & 1 
                                   \end{array} \right] \left[ \begin{array}{c}
                                                               \sin{\theta_{h0}} \cos{\phi_{h0}} \\
                                                               \sin{\theta_{h0}} \sin{\phi_{h0}} \\
                                                               \cos{\theta_{h0}}
                                                              \end{array} \right], \\
                & = & \left[ \begin{array}{c}
               \sin{(\theta_{h0} + \delta \theta_h)} \cos{(\phi_{h0} + \delta \phi_h)} \\
               \sin{(\theta_{h0} + \delta \theta_h)} \sin{(\phi_{h0} + \delta \phi_h)} \\
               \cos{(\theta_{h0} + \delta \theta_h)}
               \end{array} \right], \label{hprimedel}
\end{eqnarray}
with
\begin{eqnarray}
 \delta \theta_h = \theta_h - \theta_{h0} & = & \alpha \cos{(\phi_{c0} - \phi_{h0})},  \label{dth} \\
 \delta \phi_h  = \phi_h - \phi_{h0} & = &  \alpha \cot{(\theta_{h0})} \sin{(\phi_{c0} - \phi_{h0})}. \label{dph}
\end{eqnarray}
When $\theta_{h0} = 0$ or $\pi$, correct transformations are obtained with $\phi_{h0} - \phi_{c0} = 0$ or $\pi$ respectively, so that $\delta \phi_h$ vanishes identically. \cite{note3}

{\it In the remainder of this paper, it is deemed understood that all derivatives are calculated at oscillator equilibrium orientation} ($\theta_h = \theta_{h0}, \theta_m^h = \theta_{m0}^h$ etc.). The derivative of torque at equilibrium conditions can now be evaluated:
\begin{eqnarray}
 -  \frac{d \tau_m^\perp}{d \beta}  =  \frac{d \tau_m^\perp}{d \alpha} & = & \frac{\partial \tau_m^\perp}{\partial \theta_m^h} \frac{d \theta_m^h}{d \alpha} + \frac{\partial \tau_m^\perp}{\partial \phi_m^h} \frac{d \phi_m^h}{d \alpha}, \\
   & = & \frac{\partial \tau_m^\perp}{\partial \theta_m^h} \left( \frac{\partial \theta_m^h}{\partial \theta_{h}} \frac{d \theta_h}{d \alpha} + \frac{\partial \theta_m^h}{\partial \phi_{h}} \frac{d \phi_h}{d \alpha} \right) + \frac{\partial \tau_m^\perp}{\partial \phi_m^h} \left( \frac{\partial \phi_m^h}{\partial \theta_{h}} \frac{d \theta_h}{d \alpha} + \frac{\partial \phi_m^h}{\partial \phi_{h}} \frac{d \phi_h}{d \alpha} \right) \label{dtda1}
\end{eqnarray}
Using Eqs. (\ref{dth}) and (\ref{dph}),
\begin{eqnarray}
  -  \frac{d \tau_m^\perp}{d \beta} & = & \cos{(\phi_{c0} - \phi_{h0})} \left( \frac{\partial \tau_m^\perp}{\partial \theta_m^h} \ \frac{\partial \theta_m^h}{\partial \theta_{h}} +  \frac{\partial \tau_m^\perp}{\partial \phi_m^h} \frac{\partial \phi_m^h}{\partial \theta_{h}} \right) + \cot{(\theta_{h0})} \sin{(\phi_{c0} - \phi_{h0})} \left( \frac{\partial \tau_m^\perp}{\partial \theta_m^h} \ \frac{\partial \theta_m^h}{\partial \phi_{h}} +  \frac{\partial \tau_m^\perp}{\partial \phi_m^h} \frac{\partial \phi_m^h}{\partial \phi_{h}} \right). \label{dtauda}
\end{eqnarray}
In general, $\theta_{m}^h$ and $\phi_m^h$ may not be available as explicit functions of $\theta_h$ and $\phi_h$. The necessary derivatives at equilibrium can still be calculated in terms of the free energy density, via Eqs. (\ref{der1}) - (\ref{der4}) detailed in Appendix A. Eqs. (\ref{deltaf}), (\ref{taur2}) and (\ref{dtauda}) constitute the main result of this section. Supplemented with the equations for the determination of magnetic equilibrium ($\partial F/ \partial ( \theta_{m},\phi_m) = 0, (\partial^2F/\partial \theta_m^2) (\partial^2F/\partial \phi_m^2) - (\partial^2F/\partial \theta_m \partial \phi_m)^2 >0, \partial^2F/\partial \theta_m^2 > 0 $), the equations yield a consistent and quantitative description of CTM and TDM.

\subsection{Determination of anisotropy constants}
While Eqs. (\ref{taur2}) and (\ref{dtauda}) appear very complex at first sight, in many cases they simplify dramatically as is evident from the discussion in the next section. In any case, the anisotropy constants can be obtained by following the procedure outlined here. \\
Given TDM experimental data, we first need to assume a free energy density. Then, 
\begin{itemize}
 \item Determine the equilibrium magnetization by minimizing the free energy density.
 \item Evaluate the necessary partial derivatives using the mathematics discussed in Appendix A.
 \item Evaluate the torque derivative using Eqs. (\ref{taur2}) and (\ref{dtauda}).
 \item Evaluate the frequency shift using Eq. (\ref{deltaf}).
\item Fit the frequency shift expression thus obtained to the experimental data treating the anisotropy constants as fitting parameters.
\end{itemize}
An analytical expression for the frequency shift can be obtained in several special cases of interest. If this is not the case, one needs to follow an iterative procedure where one calculates the frequency shift numerically assuming a fixed set of anisotropy parameters, compares the calculation with the experimental data and then adjusts the assumed parameters until the numerical calculation and experimental data agree within the desired accuracy.

\section{Magnetic Field Strength Dependence}\label{MFD}

\begin{figure}[tb]
\centering
\subfloat[]{\includegraphics[width=8.5cm]{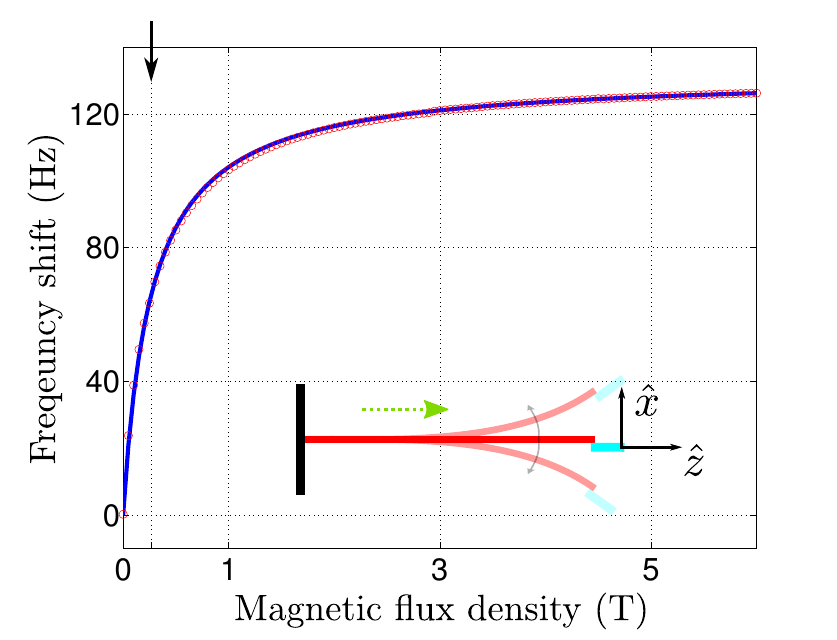}} \quad
\subfloat[]{\includegraphics[width=8.5cm]{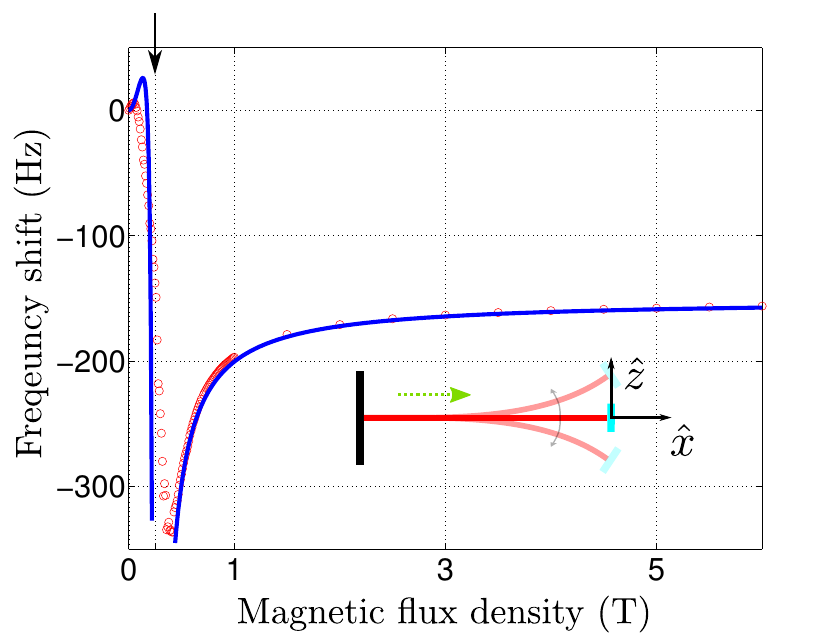}}
\caption{Frequency shift vs. applied magnetic flux density for two special cases of interest. Red open circles depict experimental data taken from Ref. \onlinecite{Weber} while the blue solid line is the frequency shift calculated from Eq. (\ref{dtauda}), using the oscillator and free energy density parameters presented in Tab. \ref{table1}. The configurations are depicted in the corresponding insets. The uniaxial easy axis is along the longer dimension of the specimen and the green dotted arrow represents the applied magnetic field. The magnitude of the uniaxial anisotropy field $B_u$ is indicated on top of the figures by a black arrow. The base resonance frequency $f_{el}$ is a few kHz.}
\label{mfsdab}
\end{figure}

The formalism developed in the previous section is now employed to calculate the frequency shift for two cases that have been investigated in literature. \cite{Stipe,Weber} We here state only the final expressions for the frequency shift, a more detailed description is given in Appendix \ref{mfsdsec}. Consider a magnetic specimen with a single easy axis along the $\hat{\pmb{z}}$ direction (a magnetic wire) so that the free energy density is given by the sum of a uniaxial anisotropy  and the Zeeman energy:
\begin{eqnarray}\label{feuni}
 F & = & K_{u} \sin^2(\theta_{m}) - \mu_0 H_{ext} M_s \left[  \sin(\theta_h) \sin(\theta_m) \cos (\phi_m - \phi_h) + \cos(\theta_h) \cos(\theta_m) \right],
\end{eqnarray}
with $K_u > 0$. The applied magnetic field is always directed along the oscillator axis unless stated otherwise.

First, the specimen shall be mounted such that its magnetic easy axis is also along the oscillator axis [see Fig.~\ref{mfsdab}(a)]. This implies $\theta_{h0} = \theta_{c0} = 0$, $\phi_{h0} = \phi_{c0}$ and the stable equilibrium solution for the magnetization direction is $\theta_{m0} = 0$ and $\phi_{m0} = \phi_{h0}$. The frequency shift is then given by
\begin{eqnarray}\label{dfuni}
   \frac{\Delta f}{f_{el}} & = & \frac{M_s V}{2 k_{el} L_e^2} \ \frac{ B_{ext} B_u}{ B_u + B_{ext}},
\end{eqnarray}
where we define $B_u = 2K_u/M_s$ as the effective anisotropy field, and $B_{ext} = \mu_0 H_{ext}$ is the applied magnetic flux density. Eq. (\ref{dfuni}) is shown as a blue solid line (using the set of parameters quoted in Tab. \ref{table1}) along with the experimental data (red open circles) from Ref. \onlinecite{Weber} in Fig.~\ref{mfsdab}(a). The agreement between experiment and theoretical model is good.

Next we consider the same sample mounted on the oscillator such that the oscillator is pointing along the $\hat{\pmb{x}}$ direction [Fig.~\ref{mfsdab} (b)]. This implies $\theta_{h0} = \theta_{c0} = \pi/2$ and $\phi_{h0} = \phi_{c0} = 0$. The equilibrium magnetization direction then is 
\begin{eqnarray}
   \phi_{m0} & = &  0, \\
   \theta_{m0} & = & \begin{cases} 
                      \sin^{-1} \left( \frac{B_{ext}}{B_u} \right)  &  B_{ext} < B_u , \\
                      \frac{\pi}{2}   & B_{ext} > B_u.
                     \end{cases} 
\end{eqnarray}
The frequency shift is accordingly obtained in the two different regimes:
\begin{eqnarray}\label{dfuni2}
   \frac{\Delta f}{f_{el}} & = &  \frac{M_s V}{2 k_{el} L_e^2} \begin{cases}
                          \frac{ B_{ext}^2 (B_u^2 - 2 B_{ext}^2) }{ B_u (B_u^2  - B_{ext}^2)} & \quad B_{ext} < B_u, \\
                          - \frac{B_{ext} B_u}{B_{ext} - B_u}  & \quad B_{ext} > B_u,
                         \end{cases}
\end{eqnarray}
and has been plotted, along with the experimental data (red open circles) from Ref. \onlinecite{Weber}, in Fig.~\ref{mfsdab}(b). We note that the frequency shift given by Eq. (\ref{dfuni2}) using a consistent free energy expression [Eq. (\ref{feuni})] for both cases is found to be in agreement with the existing literature. \cite{Stipe,Weber} We investigate some more cases of interest in Appendix \ref{mfsdsec}.

\section{High Field Limit}\label{HFL}

\begin{table}[tb]
 \begin{center}
 \begin{tabular}{|c|c|c|c|c|c|c|}
  \hline
  Set  & $L_e$ & $f_{el}$ & $k_{el}$ & $V$ & $M_s$ & $K_u$      \\ \hline
   First  &  105.4 $\mu$m & 2808.5 Hz  & 70 $\mu$N$\textrm{m}^{-1}$ & $8.3 \times 10^{-19}$ $\textrm{m}^{3}$ & 330 kA$\textrm{m}^{-1}$ & 42 $\textrm{kJ} \textrm{m}^{-3}$   \\ \hline
   Second &  105.4 $\mu$m & 2093.8 Hz  & 50 $\mu$N$\textrm{m}^{-1}$ & $7.7 \times 10^{-19}$ $\textrm{m}^{3}$ & 420 kA$\textrm{m}^{-1}$ & 52 $\textrm{kJ} \textrm{m}^{-3}$   \\
   \hline
 \end{tabular}
\caption{Oscillator and magnetic specimen parameters used for calculating frequency shift in Fig. \ref{mfsdab}(b) [second set] and all other figures (first set). Source: Ref. \onlinecite{Weber}.}
\label{table1}
\end{center}
\end{table}

Conventional torque magnetometers \cite{Chikazumi} record the torque exerted on a magnetic specimen by a large external magnetic field. In this high field limit, magnetic domains are irrelevant. In the present section, we consider the high field limit of TDM and obtain simple expressions relating the recorded frequency shift with derivatives of the free energy density. An external field much larger than the anisotropy fields in the specimen yields $\theta_m^h = \theta_h$ and $\phi_m^h = \phi_h$. Hence, the required partial derivatives are $\partial \theta_m^h/\partial \theta_h = 1, \partial \theta_m^h/\partial \phi_h = 0, \partial \phi_m^h/\partial \theta_h = 0$, and $\partial \phi_m^h/\partial \phi_h = 1$. Using these in Eq. (\ref{dtauda}), we obtain
\begin{eqnarray}\label{dtaudalphahigh}
   - \left. \frac{d \tau_m^\perp}{d \beta} \right|_{eq} & = & \cos{(\phi_{c0} - \phi_{h0})} \left. \frac{\partial \tau_m^\perp}{\partial \theta_m^h}\right|_{eq}  + \cot{(\theta_{h0})} \sin{(\phi_{c0} - \phi_{h0})}   \left. \frac{\partial \tau_m^\perp}{\partial \phi_m^h} \right|_{eq},
\end{eqnarray}
which gives the following magnetic field dependent frequency shift using Eq. (\ref{deltaf}):  
\begin{equation}\label{fshifthigh}
 \frac{\Delta f}{f_{el}} = \frac{V}{2 k_{el} L_e^2} \begin{cases}
                                                      \left. \frac{\partial^2 F^i}{\partial \theta_m^2} \right|_{eq} & \phi_{h0} = \phi_{c0}, \\
                                                      - \left( \cot{(\theta_{h0})} \left. \frac{\partial F^i}{\partial \theta_m} \right|_{eq} + \cot^2(\theta_{h0})  \left. \frac{\partial^2 F^i}{\partial \phi_m^2} \right|_{eq} \right) & \phi_{h0} = \phi_{c0} - \frac{\pi}{2}.
                                                     \end{cases}
\end{equation}
The parameters that appear in the free energy density can be extracted by fitting the frequency shift data using the above equations. The experimental configuration (viz. the magnetic field rotation plane) which is most useful will depend on the form of the free energy density.

\begin{figure}[htb]
\centering
\subfloat[]{\includegraphics[width=8.5cm]{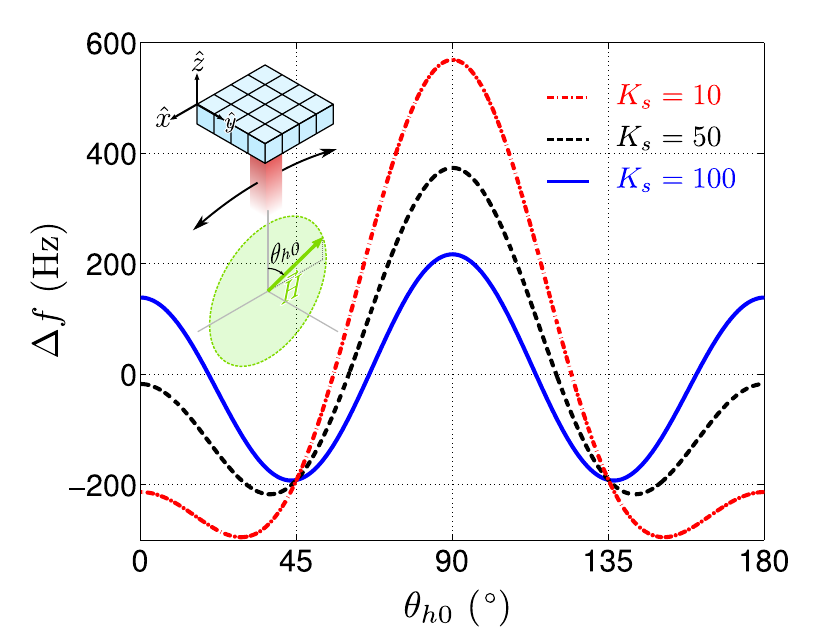}} \quad \subfloat[]{\includegraphics[width=8.5cm]{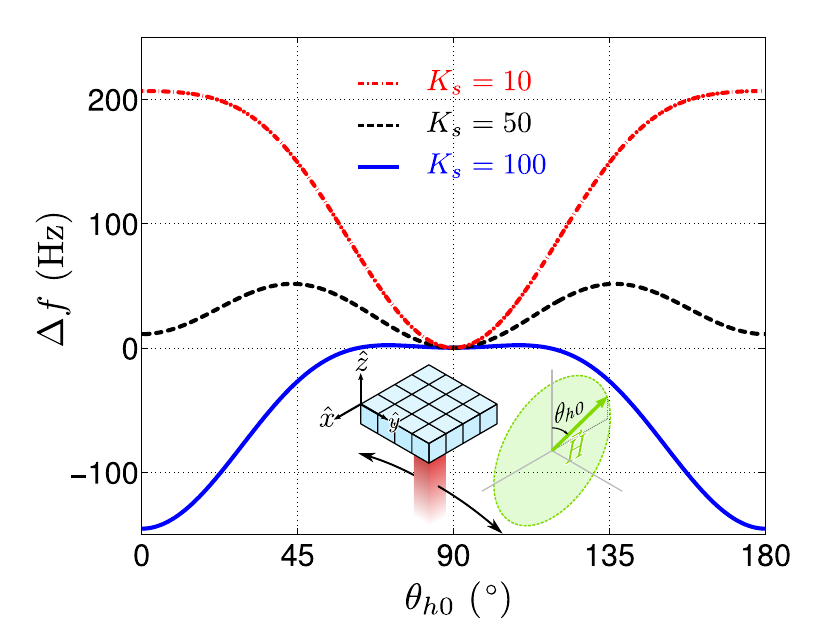}}
\caption{Frequency shift ($\Delta f$) vs. polar angle of the applied magnetic field direction ($\theta_{h0}$) for cases $\phi_{h0} = \phi_{c0} = 0$ (a) and $\phi_{h0} = \phi_{c0} - \frac{\pi}{2} = 0$ (b) from Eq. (\ref{anglerotation}). The corresponding measurement configurations are depicted in the respective insets. The cubic thin film sample (light blue) is mounted on an oscillator (red). We consider $K_1 = 47.5~\textrm{kJ m}^{-3}, K_2 = 0.75~\textrm{kJ m}^{-3}$ corresponding to magneto-crystalline anisotropy constants of Iron \cite{Chikazumi} for different values of $K_s$ (also in units of $\textrm{kJ m}^{-3}$). The qualitative shape of the curve depends upon the value of $K_s$ in relation to $K_1$. The oscillator and free energy density parameters are given in Table \ref{table1}. The base resonance frequency $f_{el}$ is about 2.8 kHz.}
\label{hf}
\end{figure}

The frequency shift for thin films with cubic magneto-crystalline anisotropy [free energy density given by Eq. (\ref{cubicfi})] can be calculated using Eq. (\ref{fshifthigh}) above:
\begin{eqnarray}\label{anglerotation}
 \frac{\Delta f}{f_{el}}  & = &  \frac{V}{2 k_{el} L_e^2} \ \begin{cases} 
                                               2 K_1 \cos(4 \theta_{h0}) - 2 K_s \cos(2 \theta_{h0}) &  \ \phi_{h0} = \phi_{c0} = 0, \\ \\
                                               K_1 \left[ 2 \cos(4 \theta_{h0}) + 3 \sin^2 (\theta_{h0}) - 4 \sin^4(\theta_{h0}) \right] - 2 K_s \cos (2 \theta_{h0}) \\ 
                                               + \frac{K_2}{2} \left[6 \sin^2 (\theta_{h0}) \cos^4 (\theta_{h0}) - 11 \sin^4 (\theta_{h0}) \cos^2 (\theta_{h0}) + \sin^6 (\theta_{h0})  \right] &  \ \phi_{h0} = \phi_{c0} = \frac{\pi}{4}, \\ \\
                                               2 \cos^2  (\theta_{h0}) \left[K_s -  (K_1 + K_2) \cos^2  (\theta_{h0}) + K_2 \cos^4  (\theta_{h0}) \right] &  \ \phi_{h0} = \phi_{c0} - \frac{\pi}{2} = 0, \\ \\
                                               \frac{\cos^2 (\theta_{h0})}{2} \left[ 6 K_1 + K_2 + 4 K_s - 10 K_1 \cos^2 (\theta_{h0}) - K_2 \cos^4 (\theta_{h0}) \right] & \phi_{h0} = \phi_{c0} - \frac{\pi}{2} = \frac{\pi}{4},
                                               \end{cases}
\end{eqnarray}
where $K_{1,2}$ characterize the cubic magneto-crystalline anisotropy, and $K_s$ parametrizes the easy plane shape anisotropy. Any of the above, but the first, configuration can be used in experiment for determining all three constants ($K_1, K_2, K_s$) in a single measurement. Fourier analysis is commonly used to isolate the contributions from different powers of the sin functions. \cite{Chikazumi} The frequency shift for the two configurations corresponding to $\phi_{h0} = 0$ is plotted in Fig.~\ref{hf}. We note that the frequency shift can get comparable to the base oscillator frequency ($f_{el}$) thereby violating our assumption of $k_m \ll k_{el}$, and necessitating use of the full resonance frequency expression Eq. (\ref{ffull}). This issue can be circumvented by relatively stiff oscillators which have higher elastic stiffness and frequency. \cite{Giessibl,Kamra}

\section{Conclusion}\label{Conclusion}

We have discussed a generic formulation for evaluating the resonance frequency shift of a mechanical oscillator mounted with a magnetic specimen as a function of the applied magnetic field. \cite{Stipe,Weber,Buchter} In addition to this frequency shift, which is measured in a TDM experiment, we also calculated a generic expression for the magnetic torque that is useful in CTM or ``DC torque magnetometry'' experiments. The latter technique, however, involves measurement of a static signal which makes it prone to noise and drift. \cite{Giessibl} Oscillators with very low $k_{el}$ are used to boost the signal which strongly limits the maximum size of the specimen that can be measured, and complicates the data analysis due to non-linearities of the oscillator. TDM, on the other hand, circumvents all of the above disadvantages, but requires the somewhat more sophisticated analysis presented here.

Equipped with the results presented herein, TDM can be a powerful technique for investigating magnetic contribution to the free energy density of a specimen. For fields large enough to saturate the magnetization along the $\pmb{H}_{ext}$ direction, we obtain relatively simple expressions for the frequency shift in terms of the free energy density [Eq. (\ref{fshifthigh})]. Given that a sensitivity large enough to investigate magnetic nano-particles via TDM has already been demonstrated, \cite{Stipe} and the progress towards simpler and cheaper experimental setups, \cite{Kamra} the calculations reported herein are expected to offer an impetus for further interest in this technique as a probe into magnetic properties of a system.

\section*{Acknowledgments}
We thank Dennis Weber, Martino Poggio and their group for sharing their experimental data. Financial support from the DFG via SPP 1538 ``Spin Caloric Transport'', Project No. GO 944/4-1, the Dutch FOM Foundation and EC Project ``Macalo'' is gratefully acknowledged.

\appendix
\section{Partial derivatives}\label{pd}
In general, it might not be possible to obtain $\theta_m^h$ and $\phi_m^h$ as closed form functions of $\theta_{h}$ and $\phi_h$. This makes the evaluation of some partial derivatives required in Eq. (\ref{dtauda}) ($\partial \theta_m^h/\partial \theta_h$ etc.) non-trivial. Here we present a method to evaluate these derivatives without having a closed form expression for $\theta_m^h$ and $\phi_m^h$.

The defining equations for $\theta_m^h,\phi_m^h$ are:
\begin{eqnarray}
 F_{\theta_m}(\theta_m^h,\phi_m^h,\theta_h,\phi_h) & \equiv & X(\theta_m^h,\phi_m^h,\theta_h,\phi_h) = 0, \\
 F_{\phi_m}(\theta_m^h,\phi_m^h,\theta_h,\phi_h) & \equiv &  Y(\theta_m^h,\phi_m^h,\theta_h,\phi_h) = 0,
\end{eqnarray}
where we have defined new functions $X$ and $Y$ for convenience. Differentiating the upper equation above w.r.t $\theta_h$:
\begin{eqnarray}
 \frac{d X}{d \theta_h}  = \frac{\partial X}{\partial \theta_h} + \frac{\partial X}{\partial \theta_m^h} \frac{\partial \theta_m^h}{\partial \theta_h} + \frac{\partial X}{\partial \phi_m^h} \frac{\partial \phi_m^h}{\partial \theta_h} & = & 0. \label{der1}
\end{eqnarray}
Similarly, by differentiating $X$ and $Y$ with respect to $\theta_h$ and $\phi_h$, we obtain:
\begin{eqnarray}
 \frac{\partial X}{\partial \phi_h} + \frac{\partial X}{\partial \theta_m^h} \frac{\partial \theta_m^h}{\partial \phi_h} + \frac{\partial X}{\partial \phi_m^h} \frac{\partial \phi_m^h}{\partial \phi_h} & = & 0, \label{der2} \\
 \frac{\partial Y}{\partial \theta_h} + \frac{\partial Y}{\partial \theta_m^h} \frac{\partial \theta_m^h}{\partial \theta_h} + \frac{\partial Y}{\partial \phi_m^h} \frac{\partial \phi_m^h}{\partial \theta_h} & = & 0, \label{der3} \\
 \frac{\partial Y}{\partial \phi_h} + \frac{\partial Y}{\partial \theta_m^h} \frac{\partial \theta_m^h}{\partial \phi_h} + \frac{\partial Y}{\partial \phi_m^h} \frac{\partial \phi_m^h}{\partial \phi_h} & = & 0. \label{der4}
\end{eqnarray}
Hence we can solve the 4 linear equations above [Eqs. (\ref{der1}) - (\ref{der4})] to obtain the 4 required derivatives $\partial \theta_m^h/\partial \theta_h$, $\partial \phi_m^h/\partial \theta_h$, $\partial \theta_m^h/\partial \phi_h$ and $\partial \phi_m^h/\partial \phi_h$ in terms of derivatives of the free energy density.

\section{Scalar products}\label{sp}
In order to evaluate the scalar products required to write Eq. (\ref{taur2}), we note the coordinate transformation between Cartesian coordinates and polar coordinates \cite{Morse2}:
\begin{eqnarray}
 \left[ \begin{array}{c}
         \hat{\pmb{r}} \\
         \hat{\pmb{\theta}} \\
         \hat{\pmb{\phi}}
        \end{array} \right] & = & \left[ \begin{array}{ccc}
                                          \sin{(\theta)} \cos{(\phi)} & \sin{(\theta)} \sin{(\phi)} & \cos{(\theta)}  \\
                                          \cos{(\theta)} \cos{(\phi)} & \cos{(\theta)} \sin{(\phi)} & -\sin{(\theta)} \\
                                          -\sin{(\phi)} & \cos{(\phi)} & 0
                                         \end{array} \right] \left[ \begin{array}{c}
                                                                     \hat{\pmb{x}} \\
                                                                     \hat{\pmb{y}} \\
                                                                     \hat{\pmb{z}}
                                                                     \end{array} \right], \\
 \tilde{P}_{\theta,\phi} & = & \tilde{S}(\theta,\phi) \ \tilde{C},
\end{eqnarray}
where the $ ~\tilde{}~ $ emphasizes that the quantity is a matrix. Therefore we obtain the following relation between the spherical unit vectors at different values of $\theta$ and $\phi$.
\begin{eqnarray}
 \tilde{P}_{\theta_{c0},\phi_{c0}} & = & \tilde{S}(\theta_{c0},\phi_{c0}) \ \tilde{C}, \\
    & = &  \left( \tilde{S}(\theta_{c0},\phi_{c0})  \tilde{S}^{-1}(\theta_{m}^h,\phi_{m}^h)  \right) \ \tilde{P}_{\theta_{m}^h,\phi_{m}^h},
\end{eqnarray}
whence we obtain:
\begin{eqnarray}
 \hat{\pmb{\phi}}_{c0} \cdot \hat{\pmb{\phi}}_{m}^h & = & \left( S(\theta_{c0},\phi_{c0})  S^{-1}(\theta_{m}^h,\phi_{m}^h) \right)_{3,3}, \\
     & = &  \cos{(\phi_{m}^h - \phi_{c0})}.  \\
 \hat{\pmb{\phi}}_{c0} \cdot \hat{\pmb{\theta}}_{m}^h & = & \left( S(\theta_{c0},\phi_{c0})  S^{-1}(\theta_{m}^h,\phi_{m}^h) \right)_{3,2}, \\
   & = & \cos{(\theta_{m}^h)} \sin{(\phi_{m}^h - \phi_{c0})} .
\end{eqnarray}

\section{Magnetic field strength dependence}\label{mfsdsec}

The formalism developed in Sec. \ref{Theory} is now applied to some special cases of interest. We start by considering a magnetic specimen with a single easy axis along the $\hat{\pmb{z}}$ direction (a magnetic wire) so that the free energy density is given by the sum of a uniaxial anisotropy  and the Zeeman energy:
\begin{eqnarray}
 F & = & K_{u} \sin^2(\theta_{m}) - \mu_0 H_{ext} M_s \left[  \sin(\theta_h) \sin(\theta_m) \cos (\phi_m - \phi_h) + \cos(\theta_h) \cos(\theta_m) \right],
\end{eqnarray}
with $K_u > 0$. {\it In the remainder of the discussion, we consider the applied magnetic field to be along the oscillator axis unless stated otherwise.}

First, the specimen shall be mounted such that its magnetic easy axis is also along the oscillator axis [see Fig.~\ref{mfsdab}(a)]. This implies $\theta_{h0} = \theta_{c0} = 0$, $\phi_{h0} = \phi_{c0}$ and the stable equilibrium solution for the magnetization direction is $\theta_{m0} = 0$ and $\phi_{m0} = \phi_{h0}$. The following expression is then obtained for the frequency shift [Fig.~\ref{mfsdab}(a)]:
\begin{eqnarray}
   \frac{\Delta f}{f_{el}} & = & \frac{M_s V}{2 k_{el} L_e^2} \ \frac{ B_{ext} B_u}{ B_u + B_{ext}}, \label{fshiftuni}
\end{eqnarray}
where we define $B_u = 2K_u/M_s$ as the effective anisotropy field, and $B_{ext} = \mu_0 H_{ext}$ is the applied magnetic flux density.

Next we consider the same sample mounted on the oscillator with a different orientation such that the oscillator is pointing along the $\hat{\pmb{x}}$ direction [Fig.~\ref{mfsdab} (b)]. This implies $\theta_{h0} = \theta_{c0} = \pi/2$ and $\phi_{h0} = \phi_{c0} = 0$. The equilibrium magnetization direction then is 
\begin{eqnarray}
   \phi_{m0} & = &  0, \\
   \theta_{m0} & = & \begin{cases} 
                      \sin^{-1} \left( \frac{B_{ext}}{B_u} \right)  &  B_{ext} < B_u , \\
                      \frac{\pi}{2}   & B_{ext} > B_u.
                     \end{cases} \label{soltheta2}
\end{eqnarray}
The frequency shift is accordingly obtained in the two different regimes:
\begin{eqnarray}
   \frac{\Delta f}{f_{el}} & = &  \frac{M_s V}{2 k_{el} L_e^2} \begin{cases}
                          \frac{ B_{ext}^2 (B_u^2 - 2 B_{ext}^2) }{ B_u (B_u^2  - B_{ext}^2)} & \quad B_{ext} < B_u, \\
                          - \frac{B_{ext} B_u}{B_{ext} - B_u}  & \quad B_{ext} > B_u.
                         \end{cases}
\end{eqnarray}
The diverging frequency shift at $B_{ext} = B_u$ renders our assumption $k_m \ll k_{el}$ invalid and requires the full expression Eq. (\ref{ffull}) for exact frequency shift calculation in a narrow window. In practice, experiments measure a large but finite response in a small applied magnetic field range.\cite{Weber} One of the advantages of this measurement scheme becomes apparent from Fig.~\ref{mfsdab}(b). The anisotropy field $B_u$ can be directly read from the plot as the field corresponding to the maximum frequency shift. The frequency shift calculated in the two cases above is found to be in agreement with the existing literature (see Fig. \ref{mfsdab}). \cite{Stipe,Weber}

\begin{figure}[tb]
\centering
\includegraphics[width=8.5cm]{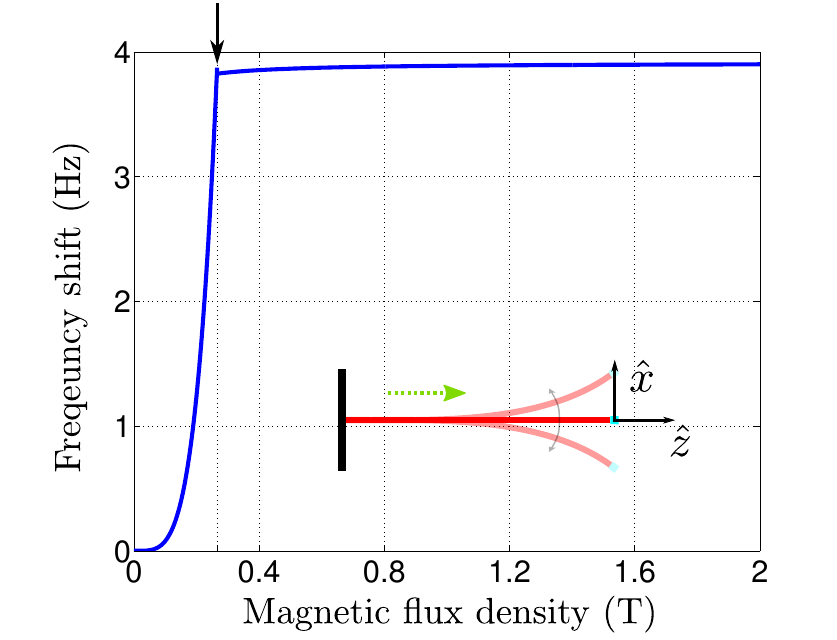}
\caption{Frequency shift vs. applied magnetic flux density. The configuration is depicted in the inset of the figure. The uniaxial easy axis is along the longer dimension of the specimen and the green dotted arrow represents the applied magnetic field. We consider a weak cubic anisotropy ( $K_c = 1~\textrm{kJ} \textrm{m}^{-3}$) in addition. The magnitude of the uniaxial anisotropy field $B_u$ is indicated on top of the figure by a black arrow. The oscillator and free energy density parameters used are quoted as the first set in Tab. \ref{table1}. The base resonance frequency $f_{el}$ is about 2.8 kHz. This measurement configuration allows for isolation of axially symmetric and polar anisotropies in a single measurement.}
\label{mfsdc}
\end{figure}

{\it Extraction of all parameters in a single measurement:} Now we consider a similar specimen as above mounted with the oscillator (and magnetic field) and oscillation direction ($\hat{\pmb{\theta}}_{c0}$) perpendicular to the easy axis. With the coordinate system used above (easy axis along $\hat{\pmb{z}}$), our assumption of the tip oscillating along $\hat{\pmb{\theta}}_{c0}$ cannot capture this configuration. Hence, we choose a different coordinate system for this case so that the easy axis is along $\hat{\pmb{y}}$ direction and the oscillator points towards $\hat{\pmb{z}}$ direction (Fig.~\ref{mfsdc}). 

The deviation of the oscillator from its equilibrium position in this configuration does not change the magnetic free energy due to the latter's axial symmetry. This implies that $\tau_m^\perp$ and hence the frequency shift should vanish for a purely uniaxial anisotropy. However, if in addition, we consider a small cubic anisotropy ($K_u \gg K_c > 0$), the total magnetic free energy density in the new coordinate system 
is given by:
\begin{eqnarray}
 F & = & \frac{K_c}{4} \left[ \sin^2(2 \theta_m) + \sin^4(\theta_m) \sin^2(2 \phi_m)  \right] - K_u \sin^2(\theta_m) \sin^2(\phi_m) \nonumber \\
   &  &  - B_{ext} M_s \left[  \sin(\theta_h) \sin(\theta_m) \cos (\phi_m - \phi_h) + \cos(\theta_h) \cos(\theta_m) \right].
\end{eqnarray}
Under the condition $ K_c \to 0$, the equilibrium magnetization orientation is given by
\begin{eqnarray}
 \theta_{m0}^h & = & \begin{cases}
                  \cos^{-1} \left( \frac{B_{ext}}{B_u} \right) & B_{ext} < B_u, \\
                  0 &  B_{ext} > B_u,
                 \end{cases} 
\end{eqnarray}
\begin{eqnarray}
  \phi_{m0}^h & = & \begin{cases}
                     \pi /2 \ \mathrm{or} \ 3 \pi / 2 & B_{ext} < B_u , \\
                     0 \ \mathrm{or} \ \pi & B_{ext} > B_u,
                    \end{cases}
\end{eqnarray}
which yields the following for the frequency shift (both values of $\phi_{m0}^h$ give the same shift):
\begin{eqnarray}\label{upluscfshift}
   \frac{\Delta f}{f_{el}} & = & \frac{M_s V}{2 k_{el} L_e^2} \begin{cases}
                          \frac{B_c B_{ext}^4}{B_u^4} &  B_{ext} < B_u, \\
                          \frac{B_{ext} B_c}{B_{ext} + B_c} &  B_{ext} > B_u,
                         \end{cases}
\end{eqnarray}
where $B_c = 2 K_c/M_s$. This configuration allows us to isolate the axial and polar dependences of the internal free energy density. Furthermore, we can deduce both parameters $K_u$ and $K_c$ from a single measurement with the magnetic field along a fixed direction. The location of the discontinuity in the slope of $\Delta f$ gives $K_u$ while the maximum frequency shift can be used to deduce $K_c$. If the sample is mounted so that the oscillator points in a direction perpendicular to the uniaxial easy axis and at an angle $\gamma$ to the cubic easy axis, the frequency shift calculated above [Eq. (\ref{upluscfshift})] is multiplied by $\cos(4 \gamma)$.

{\it Cubic magneto-crystalline anisotropy:} We now consider a {\it thick} film ($xy$ plane) specimen with strong cubic magneto-crystalline anisotropy and a weak easy plane shape anisotropy ($K_{1} \gg K_s > 0$, $K_2 > - 9 K_1$). \cite{Cullity,Chikazumi}
\begin{eqnarray}
 F^i & = & K_1 \left( m_x^2 m_y^2 + m_y^2 m_z^2 + m_z^2 m_x^2\right) + K_2 m_x^2 m_y^2 m_z^2 + K_s m_z^2, \\
   & = & \frac{K_1}{4} \left( \sin^2(2 \theta_m) + \sin^4 \theta_m \sin^2(2 \phi_m) \right) + \frac{K_2}{4} \sin^4 \theta_m \cos^2 \theta_m \sin^2 (2 \phi_m) + K_s \cos^2 \theta_m, \label{cubicfi}
\end{eqnarray}
where $m_{x,y,z}$ denote the direction cosines of the magnetization vector. We only consider the cases when the oscillator axis is along $\hat{\pmb{x}}$ and $\hat{\pmb{z}}$ (magneto-crystalline easy axes). Since the shape anisotropy has been considered weak, the equilibrium magnetization is also along the oscillator axis. \cite{note4}  
\begin{eqnarray}\label{fshiftcub}
   \frac{\Delta f}{f_{el}} & = & \frac{M_s V}{2 k_{el} L_e^2} \ \frac{ B_{ext} B_a}{ B_a + B_{ext}},
\end{eqnarray}
where $B_a$ is $B_1 + B_s$ and $B_1 - B_s$ for oscillator along $\hat{\pmb{x}}$ and $\hat{\pmb{z}}$ direction respectively, with $B_{1,s} = 2 K_{1,s}/M_s$. This implies that measurements in at least two configurations are required to obtain $K_1$ and $K_s$, while $K_2$ is not accessible to measurements along the easy axes. \cite{note5}

\begin{figure}[htb]
\centering
\includegraphics[width=8.5cm]{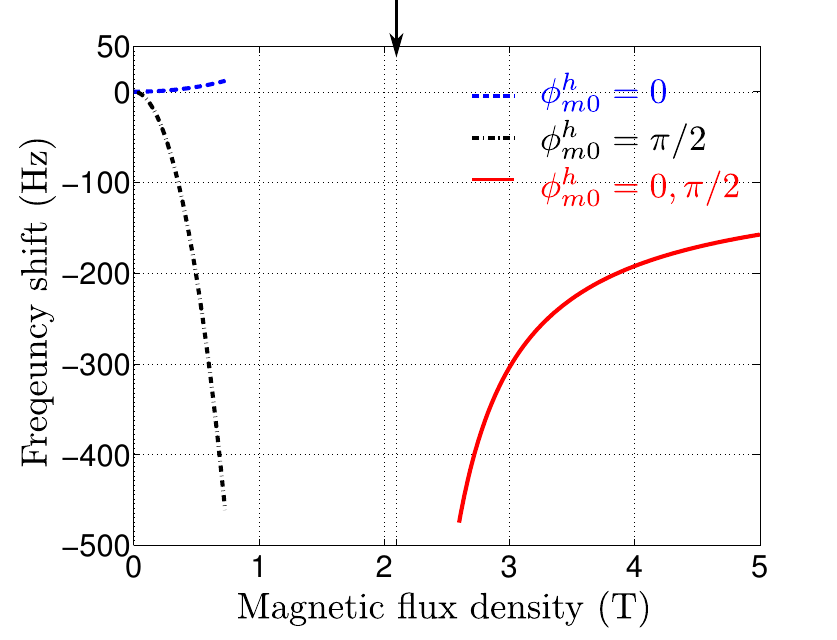}
\caption{Frequency shift vs. applied magnetic flux density for a thin film sample with cubic magneto-crystalline anisotropy. Magnetic field and oscillator axis point in the out of plane direction. We consider $V = 10^{-20} \textrm{m}^{-3} $, $K_1 = 47.2~\textrm{kJ} \textrm{m}^{-3}$ and $K_s = 1846 ~\textrm{kJ} \textrm{m}^{-3}$ corresponding to an Iron thin film \cite{Chikazumi} and oscillator parameters quoted as the first set in Tab. \ref{table1}. The base oscillator frequency $f_{el}$ is about 2.8 kHz. Energetically equivalent magnetization directions $\phi_{m0}^h = 0~\textrm{and}~\pi/2$ can easily be distinguished using low fields. There is a unique energetically favorable equilibrium orientation at high fields. The critical field separating the two regimes $B_s - B_1$ is indicated by an arrow on the top. The frequency shift close to the critical field is not shown as the expressions given in Eq. (\ref{cubK2}) are, strictly speaking, not valid in this region.}
\label{cubicK2zero}
\end{figure}

Another possibility is a magnetic thin film so that the shape anisotropy is stronger than the magneto-crystalline anisotropy ($K_s > K_1$). The case of in-plane applied magnetic field is covered by the general principle to be discussed later in the section. Here we discuss the configuration in which oscillator axis is perpendicular to the easy plane. For simplicity, we disregard the $K_2$ term in the cubic anisotropy [Eq. (\ref{cubicfi})]. The equilibrium magnetization direction is discussed in Appendix D. The frequency shift is obtained as follows.
\begin{eqnarray}\label{cubK2}
 \frac{\Delta f}{f_{el}} & = & \frac{M_s V}{2 k_{el} L_e^2} \begin{cases}
                                                                 \frac{B_{ext}^2 (B_s + B_1)^2}{(B_s + B_1)^3 - B_{ext}^2 (B_s + 7 B_1)} & \phi_{m0}^h = 0~\textrm{and}~ B_{ext} \ll B_s - B_1, \\
                                                                 - \frac{B_{ext}^2 B_s (B_s + B_1)}{B_1 \left((B_s + B_1)^2 - B_{ext}^2 \right)} & \phi_{m0}^h = \pi/2~\textrm{and}~ B_{ext} \ll B_s - B_1, \\
                                                                 - \frac{B_{ext} (B_s - B_1)}{B_{ext} - (B_s - B_1)} & B_{ext} > B_s - B_1,
                                                                \end{cases}
\end{eqnarray}
where $B_{1,s} = 2 K_{1,s}/M_s$. An analytical expression for the equilibrium magnetization, and hence the frequency shift, is not available for the middle range of magnetic flux densities (Fig.~\ref{cubicK2zero}). The orientations $\phi_{m0}^h = 0~\textrm{or}~\pi/2$ can be distinguished easily as the low field frequency shift has different signs in the two cases. One can also anticipate, on the basis of continuity, the $\phi_{m0}^h = 0$ curve in Fig.~\ref{cubicK2zero} to go to negative infinity close to $B_{ext} = B_s - B_1$. In this respect, the behavior of this curve is qualitatively similar to the case of uniaxial anisotropy considered earlier [Fig.~\ref{mfsdab} (b)]. Hence it is possible, once again, to obtain both shape and crystalline anisotropy fields in a single uni-directional measurement.

{\it Effective uniaxial anisotropy:} Eqs. (\ref{fshiftuni}) and (\ref{fshiftcub}) look identical with different anisotropy fields. This is an example of a generic principle according to which any `effective' easy axis uniaxial anisotropy field can be obtained by mounting the specimen with its easy axis along the oscillator axis. Under the mathematical conditions (which we treat as the definition of an `effective' uniaxial anisotropy):
\begin{eqnarray}
 \left. \frac{\partial^2 F^i}{\partial \phi_m \partial \theta_m} \right|_{eq} & = &  \left. \frac{\partial^2 F^i}{\partial \phi_m^2} \right|_{eq} = 0,
\end{eqnarray}
the frequency shift reduces to Eq. (\ref{fshiftcub}) with $B_a$ as the appropriate anisotropy field. In this case equilibrium magnetization direction is necessarily along the easy axis and hence the oscillator axis.

\section{Equilibrium magnetization of a thin film}\label{emotf}
We now consider the evaluation of the equilibrium magnetization direction of a thin film ($xy$ plane) with an applied magnetic field along $\hat{\pmb{z}}$. The free energy density includes a cubic magneto-crystalline anisotropy and shape anisotropy [See Eq. (\ref{cubicfi})].
\begin{eqnarray}
 F^i & = & \frac{K_1}{4} \left( \sin^2(2 \theta_m) + \sin^4 \theta_m \sin^2(2 \phi_m) \right) + K_s \cos^2 \theta_m.
\end{eqnarray}
We disregard the $K_2$ term for simplicity. We further make the following assumption: $K_s > K_1 > 0$. The equilibrium orientation of the magnetization is then given by the following equations:
\begin{eqnarray}
 K_1 (\sin(2 \theta_{m0}^h) \cos(2 \theta_{m0}^h) + \sin^3(\theta_{m0}^h) \cos(\theta_{m0}^h) \sin^2(2 \phi_{m0}^h)) -  K_s \sin(2 \theta_{m0}^h) + B_{ext} M_s sin(\theta_{m0}^h) & = & 0, \label{cond1} \\
 K_1 \sin^4(\theta_{m0}^h) \sin(4 \phi_{m0}^h) & = & 0. \label{cond2}
\end{eqnarray}
The second equation above admits $\theta_{m0}^h = 0,\pi$ or $\phi_{m0}^h = n \pi /4, n = 0,1,2 \cdots$ as possible solutions. Of these we consider only $\theta_{m0}^h = 0$ and $\phi_{m0}^h = 0 , \pi/2$ as other solutions either represent a maximum in free energy (and hence an unstable equilibrium) or solutions that are completely equivalent to the considered solutions.

Eq. (\ref{cond1}) clearly admits $\theta_{m0}^h = 0, \pi$ as a solution of which we consider only $\theta_{m0}^h = 0$ again due to energy considerations. Further $\theta_{m0}^h = 0$ does not correspond to the global minimum in energy for low fields due to shape anisotropy term. Hence we look for other solutions to the equation.
\begin{eqnarray}
 K_1 (2 \cos(\theta_{m0}^h) \cos(2 \theta_{m0}^h) + \sin^2(\theta_{m0}^h) \cos(\theta_{m0}^h) \sin^2(2 \phi_{m0}^h)) -  2 K_s \cos(\theta_{m0}^h) + B_{ext} M_s & = & 0.
\end{eqnarray}
Since we seek a solution with $\sin(\theta_{m0}^h) \neq 0$, we need $\phi_{m0}^h = 0 ~\textrm{or}~ \pi/2$ to satisfy Eq. (\ref{cond2}). For both these values of $\phi_{m0}^h$, the equation above reduces to the following:
\begin{eqnarray}
 2 K_1 \cos(\theta_{m0}^h) \cos(2 \theta_{m0}^h) -  2 K_s \cos(\theta_{m0}^h) + B_{ext} M_s & = & 0.
\end{eqnarray}
With the substitutions $\cos(\theta_{m0}^h) = x$, $K_s/K_1 = k$ and $B_{ext}/B_1 = b$, the above equation can be written as follows:
\begin{eqnarray}\label{effeq}
 2 x^3 - x (1 + k) + b & = & 0.
\end{eqnarray}
This is a cubic equation in $x$ which technically has analytic solutions, but these solutions do not offer useful insights since the expressions are rather unwieldy. We adapt an alternative approach and obtain the solution in the limit of small $b$. Clearly $x = 0$ is a solution when $b = 0$. Since the equation above is invariant with respect to the transformation $x \to - x, b \to - b$, we conclude that the Taylor expansion of $x$ in terms of $b$ will contain only odd powered terms. Hence we substitute $x = a_1 b + a_3 b^3$ in the equation above, retain terms up to $b^3$ only and obtain the following solution:
\begin{eqnarray}
 x & = & \frac{b}{1+k} + \frac{2}{1 + k} {\left( \frac{b}{1+k} \right)}^3, \\
   & = & \frac{B_{ext}}{B_s + B_1} + \frac{2 B_1}{B_s + B_1} {\left( \frac{B_{ext}}{B_s + B_1} \right)}^3.
\end{eqnarray}
The maximum value of $x$ to represent the cosine of another variable is 1. The following is true when $x = 1$ is a solution:
\begin{eqnarray}
 b & = & k - 1.
\end{eqnarray}
Since $x$ is a monotonically increasing function of $b$, we conclude that a real solution for $\theta_{m0}^h$ satisfying Eq. (\ref{effeq}) exists only for $b < k - 1$ i.e. $B_{ext} < B_s - B_1$. When this is not the case, the solution is given by $\theta_{m0}^h = 0$. Hence we have obtained equilibrium orientation of magnetization:
\begin{eqnarray}
 \theta_{m0}^h & = & \begin{cases}
                  \cos^{-1} \left( \frac{B_{ext}}{B_s + B_1} + \frac{2 B_1}{B_s + B_1} {\left( \frac{B_{ext}}{B_s + B_1} \right)}^3 \right) \approx \cos^{-1} \left( \frac{B_{ext}}{B_s + B_1} \right) & B_{ext} \ll B_s - B_1, \\
                  0 & B_{ext} > B_s - B_1,
                 \end{cases} \\
 \phi_{m0}^h & = & \begin{cases}
                0~\textrm{or}~\pi/2 & B_{ext} < B_s - B_1, \\
                \phi_{h0} = 0 & B_{ext} > B_s - B_1.
               \end{cases}
\end{eqnarray}


\begin{thebibliography}{99}
\bibitem{note1}
We consider a homogeneous magnetic field so that there is no force on the magnetic specimen due to field gradient.
\bibitem{note2}
We assume that $k_{el}$ is not modified by the slightly different equilibrium position of the oscillator in different applied magnetic fields.
\bibitem{note3}
There is an ambiguity in the value of $\phi_{h0}$ when $\theta_{h0} = 0,\pi$. This ambiguity is removed by noticing that $\delta \phi_h$ diverges for $\theta_{h0} = 0,\pi$ unless $\phi_{h0} = \phi_{c0}$. The same can also be obtained by considering the $\lim_{\epsilon \to 0^{+}} ~ \theta_{h0} = \epsilon, \pi - \epsilon$ or by substituting $\theta_{h0} = 0~\mbox{or}~\pi$ at the beginning of the transformations.
\bibitem{note4}
Strictly speaking, this will be true only for magnetic fields larger than effective shape anisotropy field.
\bibitem{note5}
$K_2$ also does not play any important role in the magnetization properties when the latter is along an easy axis.
\end{thebibliography}
\end{document}